# Electron Dynamics in Relativistic Strophotron

K.V. Ivanyan*

M.V. Lomonosov Moscow State University, Moscow 119991, Russia

**Abstract.** Relativistic strophotron is a system in which fast electrons move along a potential "trough" produced by quadrupole electric lenses. Equations of motion in the strophotron field are investigated and electron trajectories are found. It is shown, that electrons are harmonically oscillating in transverse direction and do complex motion in longitudinal direction.

## 1. Introduction

Originally the name "strophotron" had been used [1,2] for a nonrelativistic device with electrons injected and oscillating along the $x$ axis, the oscillations arising under the action of a constant electric field with a scalar potential $\Phi$ parabolically depending on $x$. The electron drift along the $z$ axis has been provided by the $y$ component of the same electric field.

In the relativistic strophotron [3-6], a relativistic electron beam is supposed to be injected at the $(x, z)$ plane under some small angle $\alpha \ll 1$ to the $z$ axis. As the motion along the $z$ axis is the main motion of the electron beam itself in this case, there is no need in the $y$ component of the electric field. In accordance with [3], we will consider here a relativistic electron beam moving in the potential "trough" produced by electric quadrupole lenses. The corresponding static potential is supposed to depend parabolically on $x$ and to be independent on $y$ and $z$.

----------------------------------

*k.ivanyan@yandex.com

Sometimes [7-11], vice versa, a crystal with electrons channeling in it is called the "solid-state strophotron." But, of course, there is a large difference between the strophotron and the crystal. The electron channeling in crystals is accompanied by many other processes such as dechanneling, bremsstrahlung, interaction with phonons, heating and damage of a crystal, absorption of radiation in the volume of a crystal, etc. All of these processes require a careful analysis in a problem of channeling, but they have no analogy in the scheme with external fields (the strophotron).

Sometimes [11] the strophotron emission is considered as analogous to the undulator emission, i.e., the emission from the system with a transversal magnetic field periodically depending on $z$. Sometimes [11] the strophotrons are not separated from the undulators at all, being called "the undulators of the second class." We will not use this name here. Of course, this is a purely terminological question. But the main idea permitting the joining of strophotrons and undulators is based on a similarity of the solutions of corresponding equations.

## 2. Electron Motion in Static Electric Field of The Strophotron

Let the scalar potential $\Phi(x)$ have the form $\Phi(x) = \Phi_0 (x/d)^2$ where $\Phi_0$ and $2d$ are the height and the width of the potential "trough." The equations of electron motion are

$$\frac{dp_x}{dt} = -\frac{2e\Phi_0}{d^2} x, \quad \frac{dp_z}{dt} = 0 \tag{1}$$

where $p_{x,z}$ are the components of the electron momentum along the x and z axes.

In accordance with the second equation (1), the longitudinal momentum and energy are conserved: $p_z = const$, $\varepsilon_z = \left(p_z^2 + m^2\right)^{1/2} = const$ (we use the system of units in which $c = 1$).

The first equation (1) can be transformed into an equation for the transversal coordinate:

$$\ddot{x} + x\Omega^2 \left(1 - \dot{x}^2\right)^{3/2} = 0, \quad \Omega^2 = \frac{2e\Phi_0}{\varepsilon_z d^2}. \tag{2}$$

If $|\dot{x}| \ll 1$, (1) for $x(t)$ turns into the equation of a harmonic oscillator having the solution

$$x(t) = x_0 \cos \Omega t + \frac{\dot{x}_0}{\Omega} \sin \Omega t \tag{3}$$

where $x_0$ and $\dot{x}_0 \approx \alpha$ are the initial transversal coordinate and speed of the electron. The condition $|\dot{x}| \ll 1$ is satisfied if $|\alpha| \ll 1$ and $|x_0|\Omega \ll 1 = 2\pi|x_0|/\lambda_0 \ll 1$ where $\lambda_0 = 2\pi c/\Omega$ is the spatial period (along the $z$ axis) of transversal oscillations of the electron. These conditions and the solution (3) will be referred to later as the "harmonic approximation."

Although $p_z = const$, the longitudinal speed $u$, depends on time $t$ because the total kinetic energy $\varepsilon_{kin} \neq const$. Under the condition $|\dot{x}| \ll 1$,

$$v_z \equiv \dot{z} = p_z/\varepsilon_{kin} \approx \frac{p_z}{\varepsilon_z}\left(1 - \frac{\dot{x}^2}{2}\right) \tag{4}$$

where

$$\frac{p_z}{\varepsilon_z} = const = \frac{\dot{z}_0}{\left(1-\dot{x}_0^2\right)^{1/2}},$$

$$\dot{z}_0 = v_0 \cos\alpha \approx 1 - \frac{1}{2\gamma^2} - \frac{\alpha^2}{2} \tag{5}$$

where $v_0$ is the total initial electron speed, $\gamma = \varepsilon_0/mc^2 \gg 1$ is the relativistic factor, and $\varepsilon_0$ is the initial electron kinetic energy.

Under the condition $|\dot{x}| \ll 1$, the anharmonicity of (2) ($\propto \dot{x}^2$) is small and can be ignored responsible for small corrections only. On the other hand, the small correction to $v_z$ (4), proportional to $\dot{x}^2$, may not be ignored because the coordinate $z$ appears often in the combination $z - t$ (see below) in which the main term of expansion (4) almost completely cancels.

Equations (3) and (4) immediately yield

$$\begin{aligned} z(t) &= \dot{z}_0\left\{\left(1 + \frac{\alpha^2 + x_0^2\Omega^2}{4}\right)t + \frac{\alpha^2 + x_0^2\Omega^2}{8\Omega}\left[\sin(2\Omega t + \varphi_0) - \sin\varphi_0\right]\right\} \\ &\approx t + \left\{-\left(\frac{1}{2\gamma^2} + \frac{\alpha^2 + x_0^2\Omega^2}{4}\right)t + \frac{\alpha^2 + x_0^2\Omega^2}{8\Omega}\left[\sin(2\Omega t + \varphi_0) - \sin\varphi_0\right]\right\} \end{aligned} \tag{6}$$

where $\varphi_0 = $ constant.

$$\sin\varphi_0 = \frac{2x_0\Omega\alpha}{\alpha^2 + x_0^2\Omega^2}, \quad \cos\varphi_0 = -\frac{\alpha^2 - x_0^2\Omega^2}{\alpha^2 + x_0^2\Omega^2}. \tag{7}$$

In the harmonic approximation (3), the amplitude of transversal oscillation of the electron is equal to $a = \left(x_0^2 + \alpha^2/\Omega^2\right)^{1/2}$. Depending on the time transversal kinetic energy, $\varepsilon_{kin}(t)$ is given by

$$\varepsilon_{\perp kin}(t) \equiv \varepsilon - \varepsilon_z = \sqrt{\varepsilon_z^2 + p_x^2} - \varepsilon_z \approx \frac{p_x^2}{2\varepsilon_z} \approx \varepsilon \frac{v_x^2}{2}. \qquad (8)$$

In the static potential $\Phi(x)$, the electron also has the potential transversal energy

$$\varepsilon_{\perp pot} = e\Phi_0 \frac{x^2}{d^2} = \varepsilon \frac{x^2 \Omega^2}{2}. \qquad (9)$$

In the harmonic approximation (3), the total transversal energy is conserved (as well as $\varepsilon_z$):

$$\varepsilon_{\perp pot}(t) \equiv \varepsilon_{\perp kin} - \varepsilon_{\perp pot} = \frac{\varepsilon}{2}\left(\alpha^2 + x_0^2 \Omega^2\right) = const. \qquad (10)$$

All the solutions described above (3) - (10) are correct under the assumption about a suddenly switching-on interaction which is applicable if $\Delta t \square 2\pi/\Omega$ where $\Delta t$ is the switching-on time or the time it takes for the electron to traverse the transitional layer $\Delta L \square c/\Delta t$. The condition $\Delta t \Omega \square 2\pi$ can be written as $\Delta L \square \lambda_0$. This inequality is satisfied if, e.g., $\Delta L \square 2-3\, cm$ and $\lambda_0 \square 10\, cm$.

### 3. Conclusion

Electron motion is described in electric field of relativistic strophotron. Electrons are oscillating in transverse direction, meanwhile they do complex motion in longitudinal direction. The strophotron system connected electron motion in both directions: transversal and longitudinal. This system can be used for creation of Free electron laser.

Under the assumption about a suddenly switching-on interaction, the electron's transversal energy jumps up at $t = 0$ to be increased by the term $\varepsilon\left(x_0^2 \Omega^2/2\right)$. However, the total energy $\varepsilon$ is conserved because at the same moment $t = 0$, the electron's longitudinal energy $\varepsilon_z$ has a jump equal to $\Delta\varepsilon_z = -\varepsilon\left(x_0^2 \Omega^2/2\right)$. This last conclusion can be confirmed by an analysis of the equations of motion in the transitional layer where the scalar potential $\Phi(x, z)$ can be taken, e.g., in the form $\Phi(x,z) = \Phi_0\left(x^2 z / d^2 \Delta L\right)$, $0 \leq z \leq \Delta L$. Writing down these equations explicitly and taking in them the limit $\Delta L \to 0$ one can check that the change of the momentum in this transitional layer is equal to $\Delta p_x = 0$ and $\Delta p_z = -\varepsilon\left(x_0^2 \Omega^2 / 2\right)$ to confirm the formulated above conclusion about the jump in $\Delta \varepsilon$. Of course, $|\Delta \varepsilon_z| \ll \varepsilon$, and this jump almost does not change the angle $\alpha$ (its change is $\sim \alpha x_0^2 \Omega^2 \ll \alpha$).